\documentclass[aps,prb,reprint,superscriptaddress,longbibliography]{revtex4-1}

\usepackage{xcolor}
\usepackage{textcomp}
\usepackage{graphicx} 
\usepackage{float}
\usepackage{epstopdf}
\usepackage{txfonts}

\usepackage[colorlinks,citecolor=blue,linkcolor=blue]{hyperref}

\newcommand{\SPEC}{SPEC, CEA, CNRS, Universit\'e Paris-Saclay, Gif-sur-Yvette, France}
\newcommand{\IAP}{Institute for Applied Physics, University of Muenster, Germany} 
\newcommand{\DEE}{Department of Electrical Engineering and ICT, University of Naples Federico II, Italy} 
\newcommand{\LabSTICC}{LabSTICC, CNRS, Universit\'e de Bretagne Occidentale, Brest, France}

\newcommand{\UMPhy}{Unit\'e Mixte de Physique, CNRS, Thales, Universit\'e Paris-Saclay, Palaiseau, France}
\newcommand{\ITEFI}{Instituto de Tecnologías Físicas y de la Información (CSIC), Madrid, Spain} 
\newcommand{\Zaragoza}{Instituto de Nanociencia y Materiales de Aragón (INMA) and Laboratorio de Microscopías Avanzadas (LMA), Universidad de Zaragoza, Zaragoza, Spain}
\newcommand{\Spintec}{Université Grenoble Alpes, CEA, CNRS, Grenoble INP, Spintec, Grenoble, France}

\begin{document}

\title{Complete identification of spin-wave eigenmodes excited by parametric pumping in YIG microdisks}

\author{T. Srivastava*}
\email{titiksha.srivastava@cea.fr}
\affiliation{\SPEC}
\author{H. Merbouche*}
\email{hugo.merbouche@uni-muenster.de}
\affiliation{\IAP}
\author{I. Ngouagnia Yemeli}
\affiliation{\SPEC}
\author{N. Beaulieu}
\affiliation{\LabSTICC}
\author{J. Ben Youssef}
\affiliation{\LabSTICC}
\author{M. Mu\~{n}oz}
\affiliation{\ITEFI}
\author{P. Che}
\affiliation{\UMPhy}
\author{P. Bortolotti}
\affiliation{\UMPhy}
\author{V. Cros}
\affiliation{\UMPhy}
\author{O. Klein}
\affiliation{\Spintec}
\author{S. Sangiao}
\affiliation{\Zaragoza}
\author{J. M. De Teresa}
\affiliation{\Zaragoza}
\author{S. O. Demokritov}
\affiliation{\IAP}
\author{V. E.  Demidov}
\affiliation{\IAP}
\author{A. Anane}
\affiliation{\UMPhy}
\author{C. Serpico}
\affiliation{\DEE}
\author{M. d'Aquino}
\affiliation{\DEE}
\author{G. de Loubens}
\email{gregoire.deloubens@cea.fr}
\affiliation{\SPEC}

\begin{abstract}

We present the parametric excitation of spin-wave modes in YIG microdisks via parallel pumping. Their spectroscopy is performed using magnetic resonance force microscopy (MRFM), while their spatial profiles are determined by micro-focus Brillouin light scattering (BLS). We observe that almost all the fundamental eigenmodes of an in-plane magnetized YIG microdisk, calculated using a micromagnetic eigenmode solver, can be excited using the parallel pumping scheme, as opposed to the transverse one. The comparison between the MRFM and BLS data on one side, and the simulations on the other side, provides the complete spectroscopic labeling of over 40 parametrically excited modes. Our findings could be promising for spin-wave-based computation schemes, in which the amplitudes of a large number of spin-wave modes have to be controlled.

\end{abstract}

\maketitle

\section{Introduction}

Novel proposals for spin-wave-based computing schemes necessitate the generation and control of multiple spin-wave (SW) modes \cite{braecher18,nakane18,hughes19,papp21,koerber22}. The most standard way to excite SW modes in a magnetic microstructure is by direct inductive coupling. There, the quasi-uniform microwave field, produced on the magnetic volume by an rf antenna, couples to the transverse dynamical component of the magnetization associated with the SW mode, with a maximal efficiency when the applied rf frequency coincides with the eigenfrequency of the mode. However, this method is not adapted to excite modes with anti-symmetric spatial profiles, as their overlap integral with the excitation field is zero \cite{naletov11}, nor short-wavelength modes, as their excitation efficiency quickly decreases with their wavevector. Yet, these two categories of modes make up a significant part of the SW k-space. In order to excite a large number of modes irrespective of their spatial profiles, parametric parallel pumping, which does not suffer from these limitations, becomes the ideal choice \cite{gurevich96}. In this case, the microwave magnetic field created by the rf antenna is aligned parallel to the static field. As a result, it does not couple to the SW modes directly. Instead, it interacts with the dynamic component of magnetization oscillating at 2$\omega$ in the static field direction, which arises due to the elliptical trajectory of magnetization precession at $\omega$. An rf field at 2$\omega$ can therefore excite SW modes at $\omega$. A quantum mechanical picture of this process is a photon generating two magnons of opposite momenta at half its frequency \cite{white63}. Since this is a nonlinear process, SWs are excited only if the amplitude of the excitation field exceeds a parametric threshold, which depends on the mode relaxation, and on the mode ellipticity. The threshold power is lower for lower relaxation rates and higher ellipticities.

Parallel pumping has been employed to generate SW modes in extended films \cite{kurebayashi11a,sandweg11,serga12,hahn13a,lauer16} and micro- and nano-waveguides \cite{mohseni20a,heinz22} of yttrium iron garnet (YIG), as well as in magnetic nanocontacts \cite{urazhdin10a}, magnetic tunnel junctions \cite{chen17a}, and micro- and nano-dots of Permalloy \cite{ulrichs11a,edwards12,guo14}. It has also been used for SW amplification \cite{braecher17b}. All these studies have been limited to a handful number of modes. The excitation and identification of many modes in an adequate system would pave the way towards simultaneous control and manipulation of a large number of SW modes for different applications in magnonics.

\begin{figure}
	\centering
        \includegraphics[width=8.5cm]{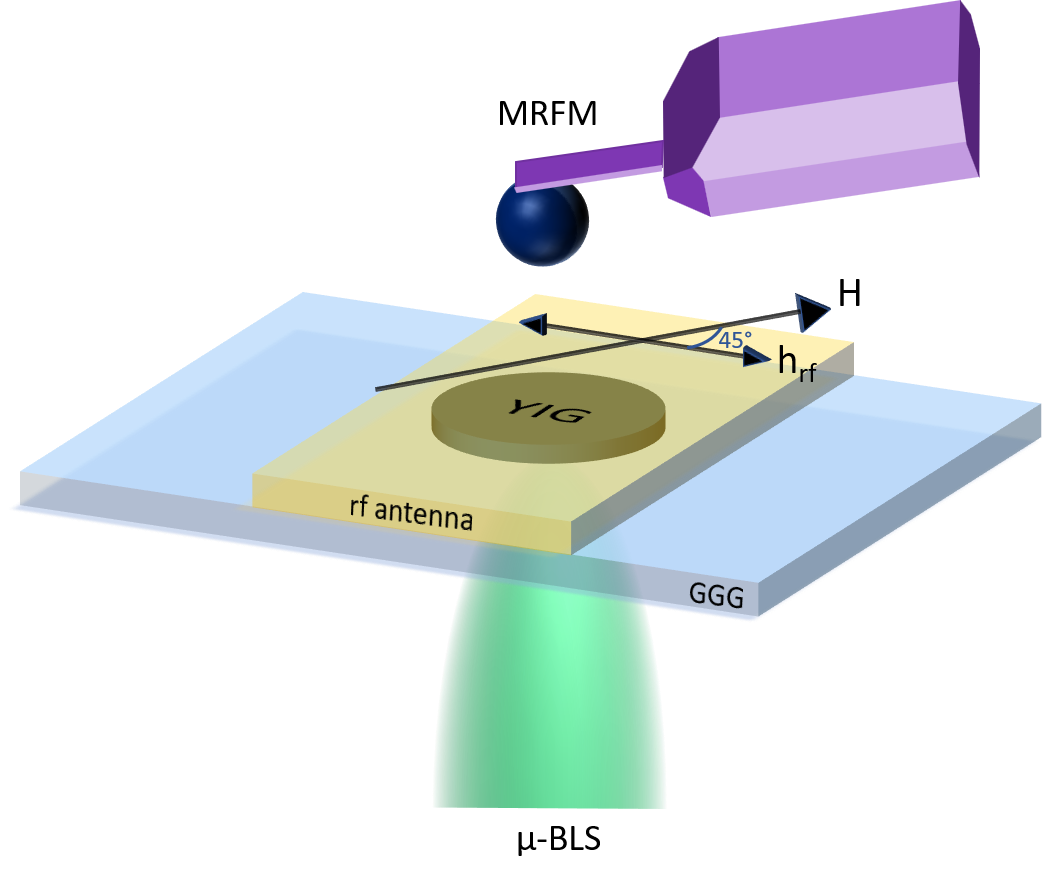}
	\caption{\textbf{Schematics of the experimental setup.} Parallel pumping of a YIG microdisk using an rf antenna deposited on top. The spectroscopy of the parametrically excited modes is achieved using a magnetic resonance force microscope (MRFM) positioned above the sample. Their spatial profiles are measured by micro-focus Brillouin light scattering (µ-BLS) using a separate experimental setup, the laser beam being focused to the bottom of the sample, through the transparent GGG substrate.}
	\label{fig:setup}
\end{figure}

 In this study, we present the excitation and identification of multiple SW modes in YIG microdisks via parametric pumping. The scheme of the experiments is shown in Fig.~\ref{fig:setup}. The SW modes are excited in YIG disks of diameters 1~µm, 3~µm and 5~µm through an integrated rf antenna and detected using a magnetic resonance force microscope (MRFM). Their spatial profiles can also be recorded using micro-focus Brillouin light scattering spectroscopy (µ-BLS). We observe that almost all the SW eigenmodes are accessible by parametric pumping. As expected, these eigenmodes become fewer in number as the size of the disk decreases. For the 3~µm disk, we label over 40 eigenmodes by comparing its MRFM parametric spectroscopy to micromagnetic simulations, and confirm the identification of as many as 10 of them through their profiles thanks to µ-BLS. Our results could be instrumental in designing basic units for unconventional computing schemes like neuromorphic computing using hyperconnected populations of a large number of eigen-excitations in a single microstructure. 

\section{Results}

\subsection{Sample}

We use 50~nm thick YIG grown on 0.5~mm thick GGG substrate by liquid phase epitaxy \cite{beaulieu18}. The characteristics of the extended film are measured by standard magnetometry and broadband FMR techniques. These yield a saturation magnetization $M_s= 140.7$~kA/m, a gyromagnetic ratio $\gamma=28.28$~GHz/T, a Gilbert damping parameter $\alpha = 7.5 \times 10^{-5}$, and a weak inhomogeneous broadening of the FMR linewidth, found to be 0.1~mT. These parameters are typical of the YIG material; the exchange constant, which has not been specifically determined on this film, is assumed to be $A = 3.7~$~pJ/m, a standard value from literature \cite{klingler15}. The YIG layer was patterned into disks of diameters 1~µm, 3~µm and 5~µm using e-beam lithography. A 220~nm thick Ti/Au antenna, of width equal to 8~µm, was then deposited on top of the disks. Injecting an rf current in the antenna generates an rf in-plane magnetic field that is orthogonal to the long axis of the antenna. 

\subsection{Parallel pumping spectroscopy}

 \begin{figure*}
	\centering
        \includegraphics[width=17.5cm]{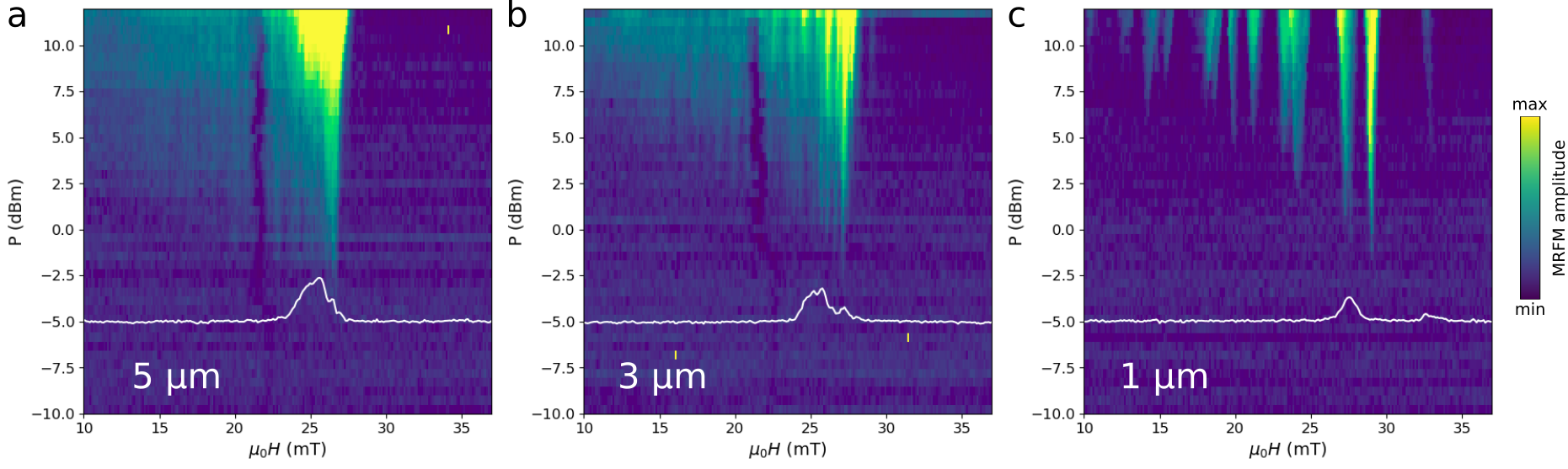}
	\caption{\textbf{MRFM parametric spectroscopy.} Intensity maps of the parametrically excited modes in the field-power coordinates excited by the microwave field of frequency 4~GHz, measured by MRFM on the 5~µm (a), 3~µm (b) and 1~µm (c) diameter YIG disks. In each panel, the continuous white curve corresponds to the direct excitation spectrum at the frequency of 2~GHz. The rf pumping field $h_\mathrm{rf}$ is at 45° from the dc field $H$.}
	\label{fig:disks}
\end{figure*}

The SW mode spectroscopy is done using MRFM. It employs a very soft cantilever, at the end of which a submicronic magnetic spherical probe made of cobalt is attached \cite{sangiao17}, to mechanically detect the magnetization dynamics in the sample placed underneath \cite{klein08}. When SWs are excited in the sample by the microwave field, the (static) longitudinal component of magnetization is reduced and so is the dipolar force on the MRFM probe, resulting in a displacement of the cantilever beam, which is detected optically. The rf excitation applied to the sample via the antenna is modulated at the mechanical resonance frequency of the cantilever to improve the quality factor and the signal-to-noise ratio. 

In these measurements, the dc magnetic field is applied in-plane at an angle of 45° with respect to the direction of the rf magnetic field, as displayed in Fig.~\ref{fig:setup}. Therefore the rf field excitation has both a transverse and a parallel component relative to the magnetization direction. The parallel pumped SW spectrum is studied for different-sized disks as a function of the applied microwave power. Figure ~\ref{fig:disks} shows the results of the MRFM parametric spectroscopy performed at a constant microwave frequency of 4~GHz for the three disks (color-coded intensity maps), together with the corresponding transverse excitation spectra measured at fixed frequency of 2~GHz and power of $-5$~dBm (continuous white curves). Only a few SW modes are detected in the latter regime. In contrast, we observe that a large number of modes can be excited by parallel pumping at 4~GHz for all the disks, in the range of applied dc field corresponding to the direct excitation of modes at 2~GHz, because parametrically excited modes are generated at half the pumping frequency. As expected, this occurs only above a minimum power level, that ranges from about $-4$~dBm for the 5~µm disk to $-2$~dBm for the 1~µm disk. The fact that the parametric threshold increases and that the density of the excited modes decreases as the lateral size decreases can be explained by geometrical confinement effects, as reported earlier\cite{guo14}.

In the following, we will mainly focus on the 3~µm disk where the SW modes are quite abundant but at the same time discernible (not too closely spaced). We perform similar measurements on this disk, this time fixing the value of the dc field to 27~mT, and scanning the parallel pumping frequency as a function of the microwave power. Fig.~\ref{fig:simu}b shows the intensity map of the parametrically excited modes in these conditions, as a function of half the pumping frequency $f_p/2$ and the rf power $P$, varied along the horizontal and vertical axes, respectively. 
We note that the threshold power increases with the frequency in a non-monotonic way, which can be explained as follows. The threshold excitation field of each mode can indeed be computed as the ratio between the relaxation rate $\omega_r(k)$ to a coupling coefficient $V(k)$, that is related to the mode ellipticity \cite{gurevich96}: the more elliptical a mode is, the larger its $V(k)$ and the lower will its threshold be. Both the terms depend non-monotonously on the wavevector $k$ and on the mode frequency. However, on a wide range, when $k$ increases, so does the mode frequency and its relaxation rate, while its ellipticity and its coupling $V(k)$ tends to decrease \cite{gurevich96,braecher17b}. This leads to the clear but not monotonous increase of the experimental threshold power with frequency seen in Fig.~\ref{fig:simu}b.

\subsection{Simulations}

\begin{figure}
	\centering
        \includegraphics[width=8.5cm]{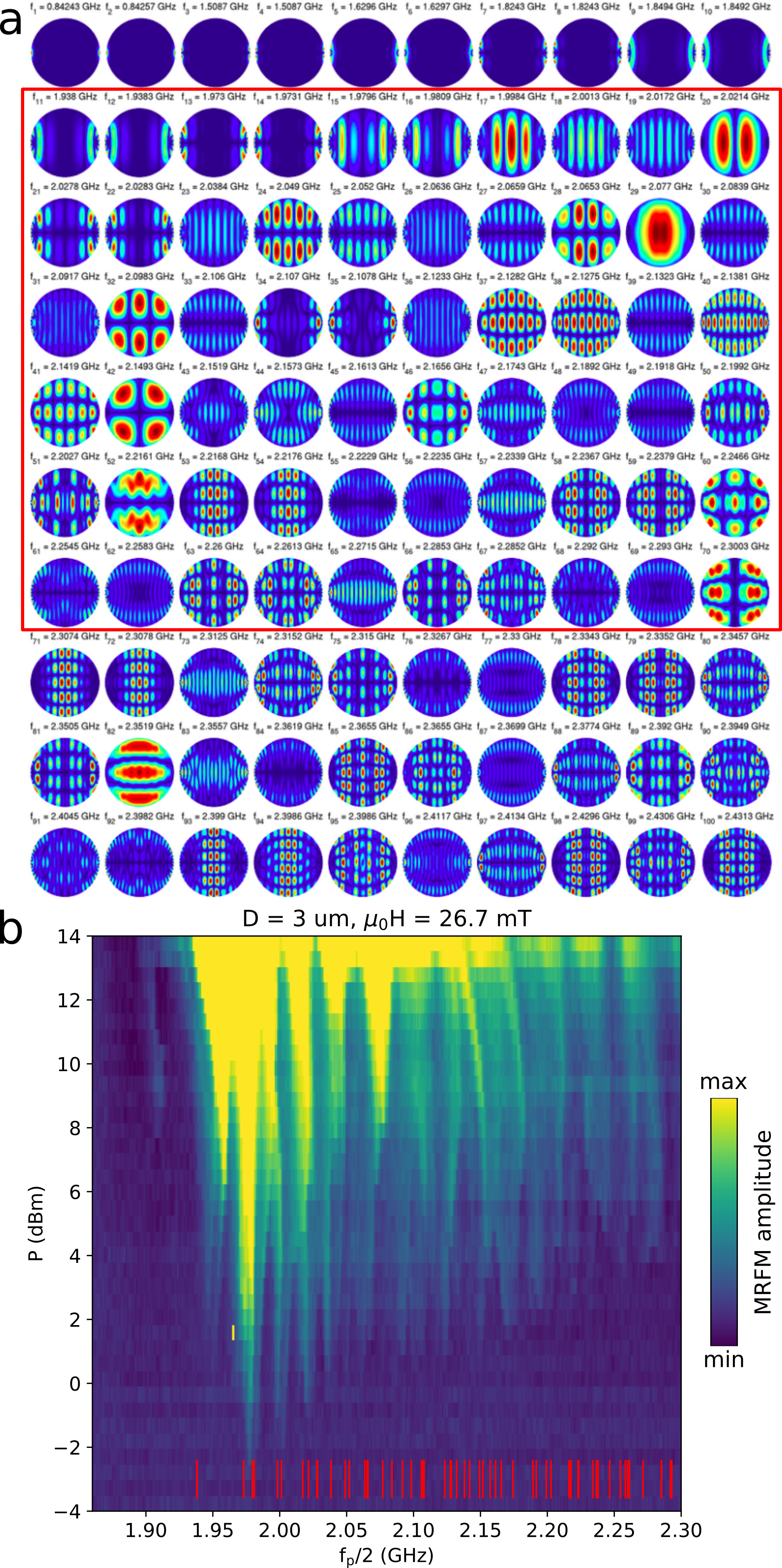}
	\caption{\textbf{Computed spin-wave eigenmodes.} (a) Computed spatial profiles of the 100 lowest frequency modes of the 3~µm diameter YIG disk, in-plane magnetized by a field of 27~mT. The color code refers to the oscillation amplitude of the local magnetization, from blue (minimum) to red (maximum). (b) Comparison between parametric spectroscopy MRFM data (color-coded intensity map) and computed eigenfrequencies (red vertical ticks) of modes 11 to 70 (surrounded by the red rectangle in panel (a)).}
	\label{fig:simu}
\end{figure}

In order to identify these parametrically excited modes, micromagnetic simulations using the eigenmode solver implemented in the micromagnetic code MaGICo \cite{magico} have been performed to calculate the SW spectrum. The magnetic ground state is first computed for the specific geometry and applied magnetic field. Once the magnetic ground state is known, the equation describing magnetization dynamics, the Landau-Lifshitz-Gilbert equation, is linearized around the ground state and small-amplitude spatial profiles of the modes are computed. This problem can be formulated as a generalized eigenvalue problem as described in ref.\cite{aquino09}. The solution of the eigenvalue problem allows to the determination of the SW spectrum of the magnetic sample under investigation.
Here, the geometry of the body, a 50~nm thick disk of 3~µm in diameter, was discretized using $300\times300\times5$ cubic cells (mesh size of $10\times10\times10$~nm$^3$), and the values of the magnetic parameters used in the simulation were those determined experimentally. As in the experimental case, the applied field lies in the plane of the disk and is set to 27~mT. The implementation of suitable matrix-free large-scale methods described in ref.\cite{aquino23} allowed the calculation of hundreds of eigenmodes for such an extended structure (353440 computational cells, eigenvalue problem size $706880\times706880$) in a few hours.

Figure~\ref{fig:simu}a displays the computed spatial profiles of the first 100 eigenmodes. The 10 lowest frequency modes (first row) correspond to edge modes, where the precession of the magnetization is strongly confined at the boundaries of the disk, in the (horizontal) direction of the applied dc field due to the demagnetizing field \cite{jersch10,guo13}. The following modes correspond to standing SW modes, which can be labelled by the number of precession lobes in the horizontal ($n_x$) and vertical ($n_y$) directions. For instance, mode 20 (second row, last column) can be labelled by $n_x=2$ and $n_y=1$, \textit{i.e.}, it is the (2,1) mode. Mode 40 (fourth row, last column) is the (11,3) mode. The most uniform mode, usually referred as the FMR mode, is mode 29, or mode (1,1).

Figure~\ref{fig:simu}b presents the comparison between the experimental spectroscopy and the computed eigenfrequencies of modes 11 to 70, shown as red ticks on top of the intensity map of the parametrically excited modes. We observe a good agreement between the computed mode frequencies and the experimental mode frequencies (at half-pumping frequencies $f_p/2$) observed at the bottom of the parametric instability regions (elongated yellow-green triangles extending downwards on the intensity map). From this comparison, it is possible to state that almost all, if not all SW eigenmodes, can be parametrically excited, irrespective of their spatial profile. Due to the high density of modes in the investigated frequency range, we will only focus on a few modes, to emphasize the good agreement noted above. The lowest-lying computed modes in Fig.~\ref{fig:simu}b are the pair of modes 11 and 12 with respective frequencies 1.938 and 1.9383~GHz, which correspond rather well to the measured parametric instability region with a threshold power of 2~dBm at around 1.95~GHz. The small disagreement of 10~MHz between the computed and measured frequencies is not unexpected, since these modes belong to the category of edge modes, whose characteristics are very sensitive to imperfections at the  periphery of the disk \cite{nembach11,guo13}, which are not taken into account in the simulations. If we move to the next parametrically excited modes, which have the lowest power threshold and have frequencies around 1.975~GHz, the comparison with computed frequencies shows that they correspond to two pairs of modes: modes 13 and 14 with respective frequencies 1.973 and 1.9731~GHz, and modes 15 and 16, at 1.9796 and 1.9809~GHz. The next excited modes in the experimental spectroscopy map are at around 2~GHz, and they correspond to mode 17 at 1.998~GHz and mode 18 at 2.001~GHz. As a matter of fact, a detailed inspection of the data shows that indeed, the parametric instability region has two nearby minima with frequencies equally spaced around 2~GHz. This good agreement between experimental and computed mode frequencies continues over the full range of investigated frequencies. We note that among the 60 modes whose frequencies have been plotted in ig.~\ref{fig:simu}b, only 44 modes have discernible frequencies and spatial profiles, a few of them being pairs of modes with very similar characteristics (e.g., pairs of modes 11 and 12, 13 and 14, 15 and 16, 21 and 22, etc.).

\subsection{Spatial profiles with µ-BLS}

\begin{figure*}
	\centering
        \includegraphics[width=16.5cm]{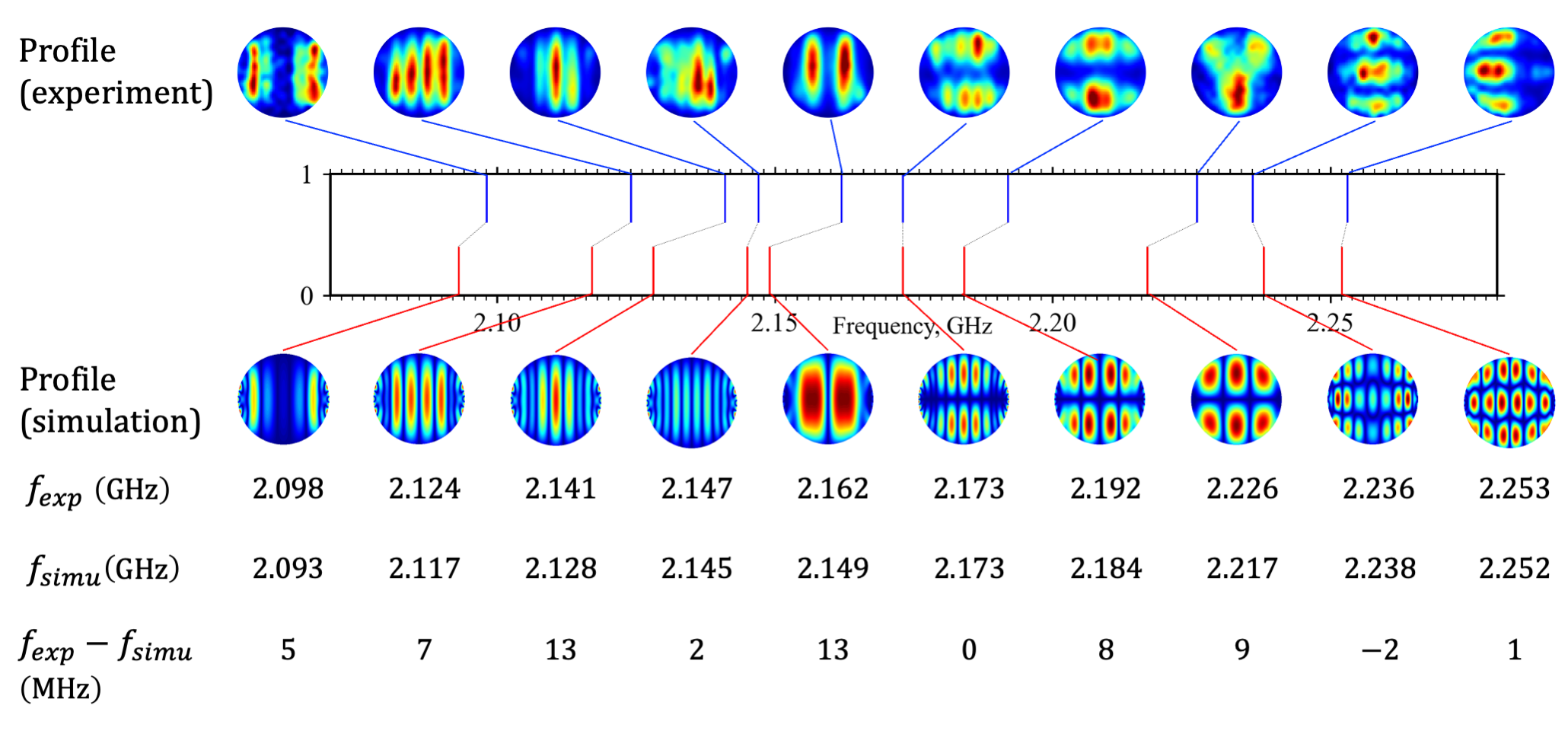}
	\caption{\textbf{BLS imaging of mode profiles.} The central graph displays the BLS detected frequencies at 30~mT for 10 modes of the 3~µm disk (blue lines) and the corresponding computed frequencies of eigenmodes (red line). These frequencies are matched (dotted dark lines) by associating the mode profiles measured in the experiment (above the graph) to the ones computed in the simulation (below). The experimental and simulated mode frequencies (in GHz) and their difference (in MHz) are given in the table. 
 }
	\label{fig:BLS}
\end{figure*}

To push further the comparison between computed SW modes and experiments, it is possible to take advantage of µ-BLS, to map the spatial profiles of dynamic magnetization in micro-structures \cite{demidov15}. A probing laser light ($\lambda=473$~nm and $P_\mathrm{laser}=0.1$~mW) is focused into a diffraction-limited spot on the surface of a similar 3~µm YIG disk (Fig.~\ref{fig:setup}) and the modulation of this probing light by the magnetization oscillations is analysed using a high-contrast optical spectrometer. The obtained signal -- the BLS intensity -- is proportional to the intensity of the magnetic oscillations at a given frequency. In this BLS measurement, the in-plane bias field is set at 30~mT. To compare the experimental mode profiles with the computed mode profiles, the micromagnetic simulations have therefore been repeated at 30~mT as well. 
To avoid nonlinear distortions of the mode profiles, known to occur when the mode amplitude increases too much, the BLS mapping of the mode profiles is performed at microwave power only slightly above threshold. By sweeping the laser spot position across the disk, a dozen of different modes are imaged, Fig.~\ref{fig:BLS} presents the comparison between the experimental and computed profiles of 10 modes. Overall, the measured profiles are in good agreement with the computed ones, taking into account the experimental spatial resolution ($\simeq 250$~nm) and the long duration of these measurements, which are subjected to experimental drifts. Similarly to the analysis performed in Fig.~\ref{fig:simu}b, we observe that the mode frequencies obtained by BLS correspond very well to the computed mode frequencies, with a mismatch that remains under 13~MHz for all modes. In particular, we observe well defined modes up to $n_x=7$ and $n_y=3$, which validates the agreement between experiment and simulations for a large number of modes.

\section{Conclusion}

Thanks to the comparison between parametric spectroscopy and mode imaging respectively performed by MRFM and BLS on one side, and micromagnetic simulations on the other side, we have successfully excited, detected and identified a large number ($>$ 40) of SW eigenmodes in a 3~µm YIG disk, where the mode density is large due to the large lateral dimensions. The computed spatial profiles provide a direct way to label those modes, using the numbers of precession nodes in the directions parallel ($n_x$) and transverse ($n_y$) to the applied magnetic field. This study opens up the possibility to perform experiments where many parametric modes are simultaneously excited while using the normal mode approach \cite{perna22,perna22a} to understand and harness the complex dynamics in the modal space of confined magnetic structures.

\section*{Acknowledgements}

This work was supported by the Horizon2020 Research Framework Programme of the European Commission under grant no. 899646 (k-NET). It is also supported by a public grant overseen by the Agence Nationale de la Recherche as part of the ``Investissements d'Avenir'' program (Labex NanoSaclay, reference: ANR-10-LABX-0035). I.N.Y. acknowledges support from the ANR grant no. ANR-18-CE24-0021 (Maestro).

\end{document}